\documentclass[%
 reprint,
superscriptaddress,
showpacs,preprintnumbers,
 amsmath,amssymb,
 aps,
prb,
]{revtex4-1}

\usepackage{graphicx}   %include figure files
\usepackage{dcolumn}% Align table columns on decimal point
\usepackage{bm}% bold math
\usepackage{hyperref}% add hypertext capabilities
\usepackage{epstopdf}

\usepackage[final]{changes}   %for final version
\begin{document}

%\preprint{APS/123-QED}

\title{Multi-photon dressing of an anharmonic superconducting many-level quantum circuit}

\author{Jochen Braum\"uller}
\author{Joel Cramer}
\author{Steffen Schl\"or}
\author{Hannes Rotzinger}
\author{Lucas Radtke}
\author{Alexander Lukashenko}
\author{Ping Yang}
\author{\added{Sebastian T. Skacel}}
\author{\added{Sebastian Probst}}
\affiliation{Physikalisches Institut, Karlsruhe Institute of Technology, 76131 Karlsruhe, Germany}
\author{Michael Marthaler}
\author{Lingzhen Guo}
\affiliation{Institut f\"ur Theoretische Festk\"orperphysik, Karlsruhe Institute of Technology, 76131 Karlsruhe, Germany}
\author{Alexey V. Ustinov}
\affiliation{Physikalisches Institut, Karlsruhe Institute of Technology, 76131 Karlsruhe, Germany}
\affiliation{National University of Science and Technology MISIS, Moscow 119049, Russia}
%\author{Gerd Sch\"on}
%\affiliation{Institut f\"ur Theoretische Festk\"orperphysik, Karlsruhe Institute of Technology, 76131 Karlsruhe, Germany}
\author{Martin Weides}
\affiliation{Physikalisches Institut, Karlsruhe Institute of Technology, 76131 Karlsruhe, Germany}


\date{\today}% It is always \today, today,

%-----------------------------------------------------------------------------

\begin{abstract}
We report on the investigation of a superconducting anharmonic multi-level circuit that is coupled to a harmonic readout resonator. We observe multi-photon transitions via virtual energy levels of our system up to the fifth excited state. The back-action of these higher-order excitations on our readout device is analyzed quantitatively and demonstrated to be in accordance with theoretical expectation. By applying a strong microwave drive we achieve multi-photon dressing \replaced{within}{of} our \replaced{anharmonic circuit}{system} which is dynamically coupled by a weak probe tone. The emerging higher-order Rabi sidebands and associated Autler-Townes splittings involving up to five levels of the investigated anharmonic circuit are observed. Experimental results are in good agreement with master equation simulations.
\end{abstract}

%-----------------------------------------------------------------------------

\maketitle
%\tableofcontents

%-----------------------------------------------------------------------------

\section{\label{sec:introduction}{Introduction}}

Superconducting quantum circuits hold great promise for the field of quantum information processing. The basic unit cell of a prospective universal quantum computer is a two-level system, referred to as a qubit. The fundamental states of superconducting qubits, being based on the Josephson effect, are mapped to the two lowest states of an oscillating system, which are isolated from higher levels by introducing an anharmonicity.
Interesting physics can be investigated when going beyond the two-level approximation. Using single-photon excitations, anharmonic oscillators have been operated as a qudit\cite{NeeleyNaturePhy08,Peterer2015}. The use of qudits was proposed to enhance the security of key distribution in the field of quantum cryptography \cite{Bruss2002,Cerf2002}. Including the third energy level, efficient and robust quantum gates have been implemented\cite{Fedorov2011}. Such three-level quantum systems provide an optimal test ground for a variety of interesting phenomena. For instance, the Autler-Townes doublet \cite{Cohen-Tannoudji1996}, induced by a dynamic Stark splitting, was observed in superconducting anharmonic circuits by coupling the first excited qubit state with one of its neighboring levels and probing the emerging splitting with a second weaker tone\cite{Baur09,SillanpaaPRL09,Novikov2013,Cho2014,Suri2013}. Furthermore, Rabi sidebands of a three-level dressed state have been investigated\cite{KoshinoPRL13}.

Dressed states in anharmonic many-level quantum circuits, as studied in this paper, are relevant to harness essential quantum properties. The underlying difficulty to predict the intrinsic behavior of many-body systems is complicated by the back-action of higher-order excitations on the dispersive readout device. Understanding the investigated many-level circuit in detail and being capable of performing a reliable and precise readout is required for prospective many-level quantum simulation\cite{Paraoanu2014}.

Here, we report on the investigation of a superconducting multi-level quantum circuit of weak anharmonicity beyond the three-level approximation. \added{In Sec.~\ref{sec:power_spec}} we spectroscopically observe multi-photon transitions via virtual energy levels up to fifth order, which is limited by the effective potential well depth of our anharmonic system. \added{A similar scheme was employed to generate single quasi-particle tunneling events in a charge qubit\cite{Graaf2013}.} In contrast to multi-photon transitions between dressed resonator-qubit states in a Jaynes-Cummings system\cite{Fink2008,Bishop2009,Kockum2013}, the reported multi-photon transitions occur in the anharmonic multi-level circuit itself.
To address and readout the anharmonic \replaced{circuit}{system}, it is embedded in a circuit quantum electrodynamic (cQED)\cite{Blais2004,Ilichev2003} environment, namely coupled to a harmonic readout resonator. \added{The back-action of higher-order excitaitons on our readout device is quantitatively investigated in Sec.~\ref{sec:power_spec_disp_shift}.}
Furthermore, by applying a microwave tone that is strong compared to the decoherence rate\added{ of our device}, we observe multi-photon dressing of states involving up to the fourth excited state of our anharmonic circuit in Sec.~\ref{sec:Rabi_AT}. Autler-Townes like avoided crossings become visible when probing the driven system with a weak microwave tone\added{, as presented in Sec.~\ref{sec:Rabi_AT}}. This demonstrates the universality of the Autler-Townes effect, which occurs when an interlevel transition is strongly and resonantly driven\cite{Cohen-Tannoudji1996}.
Observation of the associated Rabi sidebands allows to identify transitions appearing in the rotating frame, in agreement with the underlying theoretical model\added{, which is demonstrated by comparing measured data to master equation simulations.}

%-----------------------------------------------------------------------------

\section{\label{sec:setup_results}{Anharmonic many-level circuit}}

The investigated circuit is a Cooper pair box\cite{Nakamura1999} operated in the phase regime. It comprises a single Josephson junction, giving rise to the system's non-linearity, and a shunt capacitance placed in parallel. Such a circuit is commonly referred to as a transmon qubit\cite{Koch2007}. A quantum circuit of low anharmonicity allows for an extensive study of higher-order multi-photon transitions since the respective matrix elements indicating the probability of such higher order processes to occur are strongly suppressed in a more non-linear system\cite{Wuster2012}. To facilitate multi-level spectroscopy, \replaced{the amplitude of the applied microwave drive}{the applied drive power} needs to be larger than the system's anharmonicity, which in turn needs to be larger than the decoherence rate of the investigated excitations.

\subsection{\label{sec:model}{Model}}

The anharmonic quantum circuit is described by the Hamiltonian
\begin{equation}
\label{eq:cooperpair}
\hat H = 4E_C\left(\hat n\added{_{cp}} -n_g\right)^2 -E_J \cos\hat\varphi,
\end{equation}
with \replaced{$\hat n\added{_{cp}}$}{$\hat n$} being the number of excess Cooper-pairs on the transmon island, $n_g$ the offset charge and $\hat\varphi$ the Josephson phase. Shunting the Josephson junction with a large capacitance, the charging energy $E_C$ of the quantum circuit is strongly reduced, leaving the Josephson coupling with Josephson energy $E_J$ as the dominant term in the Hamiltonian. Therefore, the transmon circuit operates in the phase regime with a large ratio $E_J/E_C$ and a small associated relative anharmonicity $\alpha_r$. While this approach reduces the system's sensitivity to charge noise, it also lowers the circuit's anharmonicity\cite{Koch2007}.

The complete cQED system\cite{Blais2004} including the dispersive readout resonator is described by a generalized Jaynes-Cummings \cite{Jaynes1963} Hamiltonian, taking into account higher levels of the anharmonic circuit. In the basis of transmon states $|j\rangle$, it reads
\begin{equation}
\label{eq:Hmjc}
\hat H = \hbar \sum_j \omega_j |j\rangle \langle j| + \hbar \omega_r \hat a ^{\dagger} \hat a + \hbar \sum_{i,j} g_{ij} |i\rangle \langle j| \left( \hat a ^{\dagger} + \hat a \right),
\end{equation}
with resonance frequency $\omega_r$ of the readout resonator, eigenenergies $\omega_j$ of the anharmonic circuit and coupling matrix elements $g_{ij}$ between transitions in the anharmonic circuit and resonator. $\hat a ^{\dagger}$ ($ \hat a$) is the photon creation (annihilation) operator of the readout resonator. The Hamiltonian given in Eq.~(\ref{eq:Hmjc}) is exact, accounting for all coupling matrix elements within the considered Hilbert space limited by the potential well depth $2E_J$ (see Fig.~\ref{fig:schematic}(a)), given by the second term in Eq.~(\ref{eq:cooperpair}).

For numerical analysis of the isolated multi-level circuit, the observed transition frequencies need to be corrected for the perturbation induced by the readout resonator, being in frequency proximity to transitions in the qubit circuit. The correction can be accomplished with the Hamiltonian (\ref{eq:Hmjc}) in its canonically transformed form\cite{Koch2007} omitting terms of order $O\left(g_{i,i+1} ^2/( \omega_{i,i+1}-\omega_r)^2\right)$, taking into account only nearest neighbor coupling and applying a rotating wave approximation:
\begin{eqnarray}
\hat H ' & = & \hbar \sum_j \omega_j  |j\rangle\langle j| + \added{\hbar}\sum_{j=1} \chi_{j-1,j}|j\rangle\langle j|\nonumber\\
&&+\hbar \hat a ^{\dagger} \hat a \left(\omega_r -\chi_{01}|0\rangle\langle 0|+\sum_{j=1}\left(\chi_{j-1,j}- \chi_{j,j+1}\right)|j\rangle\langle j|\right)\nonumber\\
\label{eq:Hprime}
\end{eqnarray}
Here, \replaced{$\chi_{j,j+1}=g_{j,j+1} ^2 / \left(\omega_{j,j+1}-\omega_r\right)$}{$\chi_{i,i+1}=g_{i,i+1} ^2 / \left(\omega_{i,i+1}-\omega_r\right)$} denotes the dispersive shift induced by a transition between neighboring levels. The first line in Eq.~(\ref{eq:Hprime}) describes the change of the energy levels in the anharmonic circuit induced by the resonator while the second line gives the effective resonance frequency of the resonator dependent on the excitations being present in the many-level circuit.

\subsection{\label{sec:exp_setup}{Experimental setup}}

\begin{figure} %++++++++++++++++++++++++++++++++++++++FIGURE++++++++++++++++++++++++++++++++++++++++++
\includegraphics{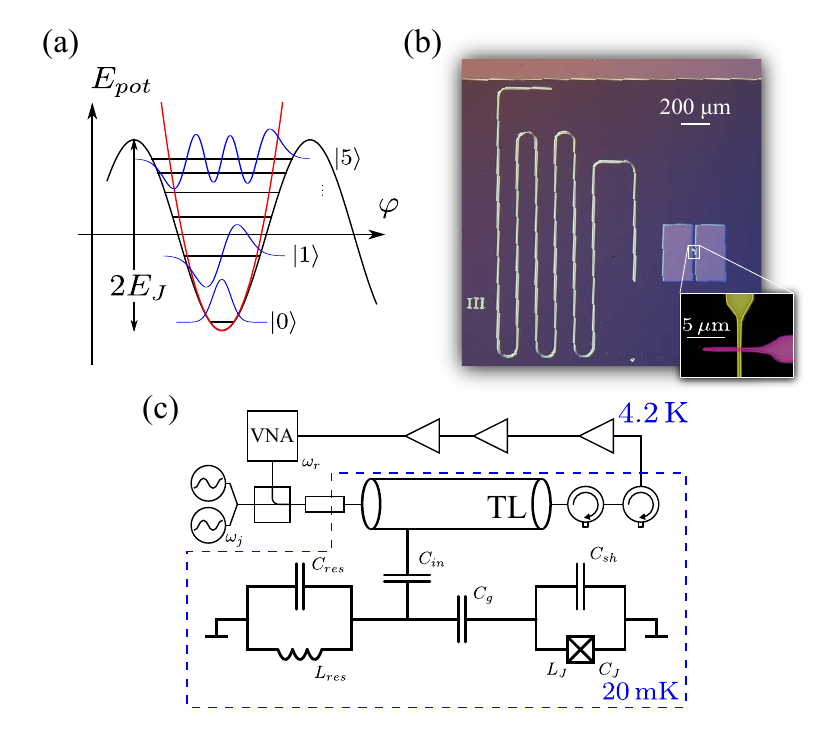}
\caption{(a) Josephson potential of depth $2E_J$ given by the cosine term in the Hamiltonian (\ref{eq:cooperpair}) with respect to the Josephson phase $\varphi$. Its approximation with a harmonic potential is indicated. (b) Optical \replaced{micrograph}{image} of the investigated chip showing part of the transmission line, the readout  resonator as well as the multi-level transmon circuit with shunt capacitance plates. The inset shows an enlarged view of the Josephson junction. (c) Schematic diagram of the cQED circuit and employed microwave setup. Resonator probe tone and anharmonic circuit manipulation tones are combined in a directional coupler and attenuated at various temperature stages of the cryostat before reaching the transmission line (TL) of the sample.}
\label{fig:schematic}
\end{figure}

The cQED system investigated in this work consists of a superconducting aluminum stripline $\lambda/2$-resonator, which is capacitively coupled to the transmon circuit, as depicted in Fig.~\ref{fig:schematic}(b). The sample is prepared in microstrip design, similar to the scheme presented in Sandberg \textit{et al.}\cite{Sandberg2013}. A backside metalization pulls the electrical field lines into the low-loss intrinsic silicon substrate of a thickness of $350\,\mathrm{\mu m}$, simplifying the layout as compared to the conventional coplanar waveguide design\cite{Weides_Transmon}. The anharmonic circuit is based on a single overlap Josephson junction shunted by two large capacitor pads arranged in plane. For simplicity of the fabrication process, we use optical lithography resulting in a Josephson junction area of about $1\,\mathrm{\mu m^2}$. Coherence is limited by the \added{relaxation rate set by the} large dielectric loss within the tunnel junction barrier \deleted{itself}\added{and at aluminum film interfaces\cite{Weides_Transmon,Weides_Trilayer}. Assuming a typical dielectric loss tangent\cite{Martinis2005} for the $\mathrm{AlO_x}$ dielectric of $\delta=2\cdot 10^{-3}$, we compute an expected relaxation time for our device due to dielectric loss to be }\added{$2\pi (\omega_{01}\delta)^{-1}=0.1\,\mathrm{\mu s}$. The Purcell limitation of the relaxation time is estimated to be $10\,\mathrm{\mu s}$, arising from the capacitive coupling of the transmon to its readout circuit.} The sample is fabricated in a two-step positive lithography process\cite{Braumueller2013} including reactive ion etching by an inductively coupled plasma. The Josephson junction consists of two magnetron sputtered aluminum electrodes, structured as overlapping thin bars in two subsequent lithography steps. The native oxide on the surface of the first aluminum layer is removed by an argon plasma cleaning step in-situ, before forming the tunnel barrier by reoxidizing in a controlled way and depositing the second electrode. With a static oxygen partial pressure of $31.5\,\mathrm{mbar}$ and an oxidation time of $60\,\mathrm{min}$, a critical current $I_c = 33\,\mathrm{nA}$ of the Josephson junction is obtained.

The quantum circuit is placed inside an aluminum box for magnetic shielding and cooled down to a base temperature of about $20\,\mathrm{mK}$ in a dilution refrigerator. The surface of the \replaced{radiation shield}{can} surrounding the base stage is coated with blackbody absorbing material to shield the sample from stray infrared light generating quasi-particles\cite{Barends2011}. Spectroscopy is performed using a vector network analyzer (VNA), measuring the transmission of continuous microwave signals applied to a transmission line coupled to the cQED system. Microwave power of the VNA as well as attenuation of the signal in the cryostat \replaced{($70\,\mathrm{dB}$)}{($-70\,\mathrm{dB}$)} is chosen such that the mean occupation of the readout resonator is at the single-photon level. Manipulation and probe signals are applied with microwave generators via the same transmission line. The outgoing signal is amplified in a high electron mobility transistor (HEMT) located at $4.2\,\mathrm{K}$ and in two amplification stages at room temperature. Two cryogenic circulators in series between the sample and the HEMT protect the quantum circuit from thermal and amplifier noise. The schematic experimental wiring and the circuit diagram of the sample are depicted in Fig.~\ref{fig:schematic}(c).

%-----------------------------------------------------------------------------

\section{\label{sec:power_spec}{Power spectroscopy}}

\begin{figure*} %++++++++++++++++++++++++++++++++++++++FIGURE++++++++++++++++++++++++++++++++++++++++++
\includegraphics{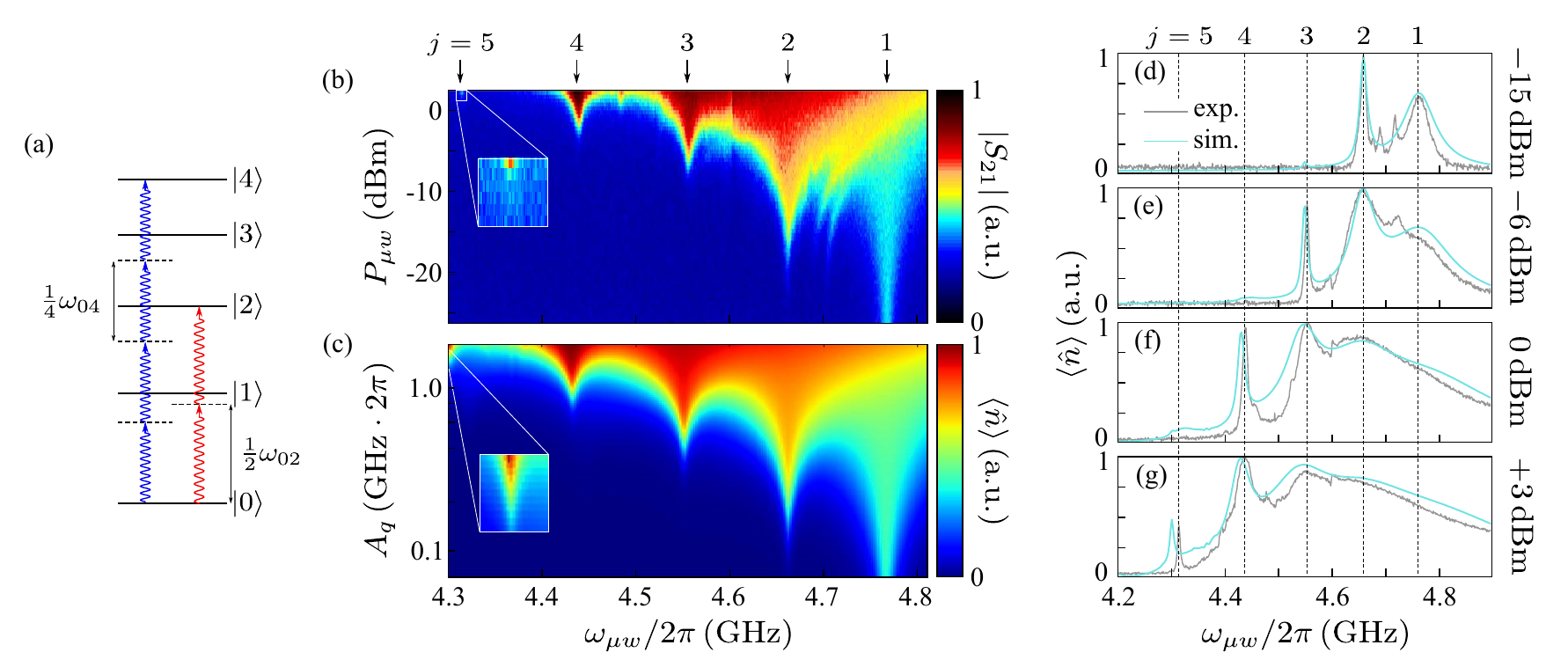}
\caption{(a) Energy diagram illustrating the observed multi-photon transitions. (b) Power spectrum of the multi-level circuit showing the transmission magnitude \replaced{$|S_{21}|$}{$S_{21}$} of the readout resonator corresponding to the excitation number \replaced{$\langle \hat n\rangle$}{$\langle n\rangle$} of the many-level circuit. By increasing the drive power $P_{\mu w}$, multi-photon transitions $1/j\,\left( |0\rangle \leftrightarrow |j\rangle\right)$ up to fifth order are observable at certain transition frequencies $\omega_{\mu w}$. (c) Respective master equation simulation. (d)-(g) Detailed comparison of experimental and simulated data for horizontal cuts of the power spectrum at distinct drive powers. Power values $P_{\mu w}$ are given without attenuation in the cryostat. From fitting experimental and simulated data, the conversion factor of the power scale $P_{\mu w}$ into an amplitude $A_q$ that directly couples to the multi-level circuit is found to be approximately $A_q=1.4\cdot 10^{P_{\mu w}/20\,\mathrm{dBm}} \,[\mathrm{GHz}\cdot 2\pi ]$. }
\label{fig:power_spectrum}
\end{figure*}

%\subsection{\label{sec:power_spec_exp}{Experimental data}}

The properties of the anharmonic multi-level circuit are experimentally investigated by applying a continuous microwave drive tone of frequency $\omega_{\mu w}$ and power $P_{\mu w}$ \added{to excite transitions present in the circuit. The level population $\langle\hat n \rangle$\added{, $\hat n=\sum_j j|j\rangle \langle j|$,} of the anharmonic circuit is measured via the dispersive shift \replaced{$\chi_{ij}$}{$\chi$} in resonance frequency of the readout resonator. For efficient data acquisition, the transmission magnitude $|S_{21}|$ is measured at a frequency close to resonance of the readout device at $\omega_r^m/2\pi=5.827\,\mathrm{GHz}$, which allows to infer the dispersive resonator shift and therefore the level population of the anharmonic circuit.} \deleted{The dispersive resonator shift $\chi$ indicating the level population $\langle\hat n \rangle$ of the anharmonic circuit corresponds to the transmission magnitude $|S_{21}|$ measured at a frequency close to resonance.}\added{After presenting measured power spectroscopy data we extract the bare frequencies of the complete cQED system by means of Eq.~(\ref{eq:Hprime}) to reconstruct the parameters of the isolated many-level circuit. The necessary dispersive shifts $\chi_{ij}$ to conduct this analysis are obtained in Sec.~\ref{sec:power_spec_disp_shift} by comparing measured data with theoretical expectation. The extracted parameters of the bare many-level circuit are the basis for the numerical investigation presented in Sec.~\ref{sec:power_spec_sim}.}

The experimentally obtained power spectrum of the investigated circuit is shown in Fig.~\ref{fig:power_spectrum}(b). For small drive powers, only the fundamental qubit transition to the first excited state $|1\rangle$ is visible ($j=1$). Since the multi-level quantum circuit is cooled down to its ground state, only transitions from the ground state $|0\rangle$ to any higher level $|j\rangle$ of the system are observable. The matrix elements accounting for direct coupling between non-neighboring levels are strongly suppressed due to parity selection rules. However, transitions between levels $|0\rangle$ and $|j\rangle$, $j=2..5$, occur as multi-photon transitions involving $j$ photons via $j-1$ virtual intermediate energy levels. These $j$-photon transitions are observable in power spectroscopy (Fig.~\ref{fig:power_spectrum}(b)) as transition peaks at distinct frequencies and are denoted as $1/j\,\left( |0\rangle \leftrightarrow |j\rangle \right)$ throughout this text. Multi-photon transitions are indicated in Fig.~\ref{fig:power_spectrum}(a), showing the energy diagram of the circuit.
With increasing drive power, the line width strongly increases from its intrinsic value. This power broadening occurs as more states contribute to achieve level saturation.

\replaced{The fundamental transition $|0\rangle \leftrightarrow |1\rangle$ appears at the highest drive frequency and is solely visible at lowest drive powers. Since the anharmonicity of the circuit is negative, as indicated in Fig.~\ref{fig:power_spectrum}(a), the next order transition $1/2\,\left( |0\rangle \leftrightarrow |2\rangle\right)$ occurs at a lower frequency. The absolute value of the line separation equals $|\alpha|/2h=E_C/2h=96\,\mathrm{MHz}$, with $\alpha =\hbar (\omega_{12}-\omega_{01})$ being the absolute anharmonicity of the many-level circuit\cite{Koch2007}.}{The fundamental transition $|0\rangle \leftrightarrow |1\rangle$ and the next order transition $1/2\,\left( |0\rangle \leftrightarrow |2\rangle\right)$ appear at highest drive frequencies, since the anharmonicity of the circuit is negative.} With the measured transition frequencies $\omega_{01} ^m/2\pi=4.766\,\mathrm{GHz}$, $\omega_{12} ^m/2\pi=4.558\,\mathrm{GHz}$ as extracted from power spectroscopy (Fig.~\ref{fig:power_spectrum}(b)), the transition frequencies $\omega_{i,i+1}$ of the isolated anharmonic circuit can be found via the relations
\begin{eqnarray}
\omega_{01} ^m & = & \omega_{01} + \chi_{01},\nonumber\\
\omega_{12} ^m & = & \omega_{12} + \chi_{12} - \chi_{01},
\label{eq:omegacorr}
\end{eqnarray}
\added{applying Eq.~(\ref{eq:Hprime}). \replaced{$\chi_{j,j+1}$}{$\chi_{i,i+1}$} are calculated according to the definition, employing the coupling matrix elements \replaced{$g_{j,j+1}$}{$g_{i,i+1}$} as extracted in the analysis presented in Sec.~\ref{sec:power_spec_disp_shift}.}
Likewise, the resonance frequency $\omega_r/2\pi=5.814\,\mathrm{GHz}$ of the isolated readout resonator is extracted, employing the undressed transition frequencies \replaced{$\omega_{j,j+1}$}{$\omega_{i,i+1}$} and the unperturbed resonator frequency $\omega_r$ recursively in the definition of \replaced{$\chi_{j,j+1}$}{$\chi_{i,i+1}$}.

\begin{table} %++++++++++++++++++++++++++++++++++++++TABLE++++++++++++++++++++++++++++++++++++++++++
\begin{center}
\caption{\added{Parameters of the complete cQED system and the isolated many-level circuit, extracted according to Eq.~(\ref{eq:omegacorr}). It is evident from comparing measured and bare frequencies that levels are repelled due to the strong coupling in the Jaynes-Cummings system.} \added{The shift in resonance frequency $\omega_r$ of the increasingly decoupled readout resonator towards lower frequencies can be validated by measuring its resonance frequency for a strongly increased probe power (not shown here).}}
\label{tab:parameters}
\vspace*{1mm}
\begin{tabular}{p{1cm}p{2.5cm}p{2.3cm}}\toprule
& measured $\omega^m/2\pi\,(\mathrm{GHz})$ & bare $\omega/2\pi\,(\mathrm{GHz})$\\
\hline
$\omega_r$ & 5.827 & 5.814\\
$\omega_{01}$ & 4.766 & 4.779\\
$\omega_{12}$ & 4.558 & 4.565\\
\hline
\hline
\end{tabular}
\\
\vspace*{1mm}
$\alpha_r=-4.5\%$, $E_C=h\cdot 0.19\,\mathrm{GHz}$, $E_J/E_C=83$
\end{center}
\end{table}

With the transition frequencies $\omega_{01}/2\pi=4.779\,\mathrm{GHz}$ and $\frac 1 2 \omega_{02}/2\pi=4.565\,\mathrm{GHz}$ of the isolated many-level circuit, the relative anharmonicity $\alpha_r$ is extracted to be $\alpha_r = \omega_{02}/\omega_{01}-2 = -4.5\%$. Writing the Hamiltonian (\ref{eq:cooperpair}) in phase basis allows to extract its exact eigenenergies and we obtain a ratio $E_J/E_C = 83$ and thereby the complete energy spectrum of the investigated multi-level circuit. \added{Table \ref{tab:parameters} summarizes the relevant parameters.}

The given level identification is confirmed by observing the $|1\rangle\leftrightarrow|2\rangle$ transition in a separate two-tone experiment, while strongly pumping the first excited state $|1\rangle$.

Figure~\ref{fig:schematic}(a) illustrates the deviation of the anharmonic Josephson potential, given by the second term in Eq.~(\ref{eq:cooperpair}), from the parabolic potential of a harmonic oscillator. The depicted eigenenergies of the anharmonic circuit are found by diagonalization of the Hamiltonian (\ref{eq:cooperpair}) after writing the Josephson coupling term as a tunneling operator $-E_J/2\,\sum_n |n\rangle\langle n+1|+|n+1\rangle\langle n|$. The obtained ground state energy $E_0=h\cdot 2.4\,\mathrm{GHz}$, specified relative to the minimum of the cosine potential set by $-E_J=-h\cdot\,16.1\,\mathrm{GHz}$, equals the vacuum energy $\frac 1 2 \hbar\omega_{01}$ for the state $|0\rangle$ of a harmonic oscillator with energy splitting $\omega_{01}$. While multi-photon transitions are observable up to the fifth excited level, which still is a bound state well within the cosine potential, the sixth excited level is roughly $190\,\mathrm{mK}$ below the energy barrier and therefore excitations of this state are likely to leak to a delocalized state.

The two power dependent features visible in measured data (Fig.~\ref{fig:power_spectrum}(b)) at $4.7\,\mathrm{GHz}$ are not intrinsic to the investigated quantum circuit but rather correspond to chip modes originating from neighboring circuits on the investigated sample. These features do not shift in frequency between different cool-downs, as opposed to a shift in transition frequencies of the investigated circuit caused by an aging effect of the Josephson junction. The observed power dependence might be due to a coupling of these modes to the broadened multi-photon transitions.

\subsection{\label{sec:power_spec_disp_shift}{Quantitative investigation of the dispersive resonator shift}}

\begin{figure} %++++++++++++++++++++++++++++++++++++++FIGURE++++++++++++++++++++++++++++++++++++++++++
\includegraphics{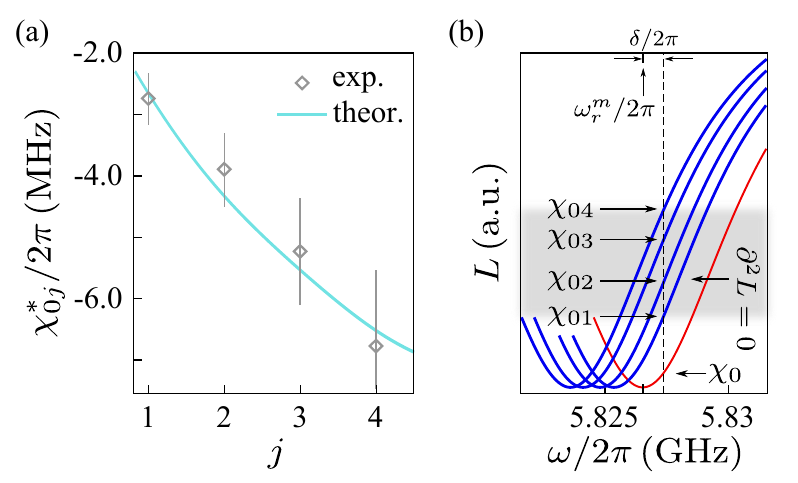}
\caption{(a) Comparison of the measured saturated dispersive shift \replaced{$\chi_{0j}^*$}{$\chi_{0j}$} induced by $j$-photon transitions to the theoretical expectation for $j=1..4$. The blue line is a guide to the eye of calculated quantized values. Data points in gray are experimental values extracted from power spectroscopy shown in Fig.~\ref{fig:power_spectrum}(b) and corrected for the non-linearity introduced by the measurement technique. Good agreement is achieved for a coupling constant \replaced{$g_{01}/2\pi=115\,\mathrm{MHz}$}{$g_{01}=2\pi\cdot\,115\,\mathrm{MHz}$}. (b) \replaced{Illustration of the readout technique: Lorentzians $L$ indicate the squared transmission magnitude $|S_{21}(\omega)|^2$ data obtained by different dispersive shifts of the readout resonator. With no excitation being present in the anharmonic circuit, the resonator is subject to the offset shift $\chi_0$ (red). In blue, dispersive resonator shifts corresponding to the observed multi-photon transitions are indicated. The depicted Lorentzians are reconstructed from measured data.}{Lorentzian fit $L$ (red) to the squared transmission magnitude $|S_{21}(\omega)|^2$ of the readout resonator. With no excitation being present in the anharmonic circuit, the resonator is subject to the offset shift $\chi_0$. In blue, resonator shifts corresponding to the observed multi-photon transitions are indicated.} The readout frequency $\omega_r ^m+\delta$ (dashed line) is chosen to be close to the linear point of the Lorentzian where \replaced{$\partial ^2_\omega L=0$}{$\partial ^2 L=0$}.}
\label{fig:dispshift_quant}
\end{figure}

As mentioned earlier, dispersive readout of the anharmonic many-level circuit is performed via the induced back-action on its readout resonator, namely the dispersive shift \replaced{$\chi_{ij}$}{$\chi$} of its resonance frequency. According to the Jaynes-Cummings model\cite{Blais2004}, resonator levels are dressed dependent on the excitation number $\langle\hat n\rangle = \sum_i i|i\rangle\langle i|$ of the many-level circuit, such that \replaced{$\chi_{ij}\propto\langle\hat n\rangle$}{$\chi\propto\langle\hat n\rangle$}.
For investigating a two-level qubit system, it is sufficient to observe a relative change in the resonator's frequency $\omega_r$ to infer the qubit state. Going beyond the two-level approximation however requires to consider absolute values of \replaced{$\chi_{ij}$}{$\chi$} in order to extract $\langle\hat n\rangle$ of the many-level circuit. To reduce measurement time, the dispersive shift \replaced{$\chi_{ij}$}{$\chi$} is detected by measuring the transmission magnitude \replaced{$|S_{21}(\omega_r^m+\delta)|$}{$|S_{21}(\omega_r+\delta)|$} at a frequency detuned by $\delta$ from resonance. Since there is no linear proportionality \added{in general}, \replaced{$|S_{21}(\omega_r^m+\delta)|\not\propto\chi_{ij}$,}{$|S_{21}(\omega_r+\delta)|\not\propto\chi$} \deleted{in general,} corrections to extract \replaced{$\chi_{ij}$}{$\chi$} from this measurement technique are necessary.

Figure~\ref{fig:dispshift_quant}(a) quantitatively shows the \added{relative} dispersive resonator shift \replaced{$\chi_{0j}^*$}{$\chi_{0j}$} induced by multi-photon transitions of $j$-th order. The theory line results from the saturated dispersive shifts \replaced{$\chi_{0j}^*$}{$\chi_{0j}$}, for the situation where both the driven level $|j\rangle$ and the ground state $|0\rangle$ are populated equally, $|n\rangle=\frac 1 2 |0\rangle +\frac 1 2 |j\rangle$. According to the second line of Eq.~(\ref{eq:Hprime}),\added{ the relative dispersive shift $\chi_{0j}^*$ takes the form}
\begin{eqnarray}
\chi_{0j}^* & = & \frac 1 2 \left( \frac{g_{j-1,j} ^2}{\omega_{j-1,j}-\omega_r} - \frac{g_{01} ^2}{\omega_{01}-\omega_r} - \frac{g_{j,j+1} ^2}{\omega_{j,j+1}-\omega_r} \right) -\chi_{0}\nonumber\\
& = & \frac 1 2 \left( \frac{g_{j-1,j} ^2}{\omega_{j-1,j}-\omega_r} - \frac{g_{j,j+1} ^2}{\omega_{j,j+1}-\omega_r} \right)\nonumber\\
& = & \frac 1 2 \left(\chi_{j-1,j}-\chi_{j,j+1}\right)
\label{eq:chi_ij}
\end{eqnarray}
for $j>0$, employing a coupling constant \replaced{$g_{01}/2\pi=115\,\mathrm{MHz}$}{$g_{01}=2\pi\cdot\,115\,\mathrm{MHz}$}. Higher-order coupling matrix elements $g_{i,i+1}$ are calculated by solving the Hamiltonian (\ref{eq:cooperpair}) in charge basis and evaluating the respective matrix elements. \added{As the readout resonator is subject to an offset shift $\chi_{0}=-\chi_{01}=-g_{01}^2/(\omega_{01}-\omega_r)$, induced by the quantum vacuum energy of the multi-level circuit, $\chi_{0j}^*$ is given relative to $\chi_0$ to reproduce measured data.}\deleted{The offset shift \replaced{$\chi_{0}=-\chi_{01}=-g_{01}^2/(\omega_{01}-\omega_r)$}{$\chi_{0}=-g_{01}^2/(\omega_{01}-\omega_r)$} of the resonator induced by the quantum vacuum energy of the multi-level circuit is subtracted, respectively.}

Data points in gray are the experimentally measured saturated dispersive shifts \replaced{$\chi_{0j}^*$}{$\chi_{0j}$}, corrected for the non-linearity induced by the described measurement technique. This is realized by fitting the squared linear transmission magnitude $|S_{21}(\omega)|^2$ of the readout resonator to a Lorentzian. Its associated decay rate $\kappa/2\pi=8.7\,\mathrm{MHz}$ is previously found by a complex circle fit\cite{Probst2015} of measured transmission magnitude data. To obtain the given data points, we extract the saturated transmission magnitudes $|S_{21}^j(\omega_r ^m+\delta)|$ of the observed $j$-photon transitions, corresponding to equal level population, from power spectroscopy data presented in Fig.~\ref{fig:power_spectrum}. Error bars are \replaced{Gaussian}{Gauss} errors taking into account fit errors in the measured transmission magnitudes $|S_{21}|$, the decay rate $\kappa$, the base line and minimum of the fitted Lorentzian as well as the readout detuning $\delta$ from resonance.
Good agreement between experimental data and the theoretical model is demonstrated.

It is interesting to note that comparing the experimentally obtained saturated dispersive shifts with the theoretical expectation provides an opportunity to extract the fundamental coupling constant $g_{01}$ between \added{the} many-level circuit and \added{the} readout resonator, since $\chi_{01}\propto g_{01} ^2$.

Selecting a proper detuning $\delta$ allows for quasi-linear operation as illustrated in Fig.~\ref{fig:dispshift_quant}(b). Since the linear point of the resonator's Lorentzian $L$, where \replaced{$\partial ^2_\omega L=0$}{$\partial ^2 L=0$}, is well within the measurement region, required corrections to translate the measured transmission magnitude $|S_{21}(\omega_r ^m+\delta)|$ into the dispersive shift \replaced{$\chi_{ij}$}{$\chi$} are small.

The drive power in the experiment was kept below the threshold for achieving saturation of the $\frac 1 5 |0\rangle\leftrightarrow|5\rangle$ transition due to experimental setup constraints.

\subsection{\label{sec:power_spec_sim}{Simulation}}

Validation of the underlying theoretical model is carried out via the comparison of experimental data to a numerical simulation as shown in Fig.~\ref{fig:power_spectrum}(c). It is based on a master equation solver provided by the QuTIP package\cite{Johansson2011,Johansson2012}. We are plotting the expectation value $\langle \hat n\rangle$ of the occupation number operator of the multi-level circuit in a color scale, which is linearly proportional to the energy of the readout resonator and therefore to its dispersive shift \replaced{$\chi_{ij}$}{$\chi$}. Parameters are the first two transition frequencies $\omega_{01}$, $\frac 1 2 \omega_{02}$ of the isolated multi-level circuit and the resonance frequency $\omega_r$ of the bare readout resonator together with its decay rate $\kappa$ as determined in the previous section. The relaxation rate $T_1=56\,\mathrm{ns}$ of the fundamental transition $|0\rangle\leftrightarrow|1\rangle$ is obtained in a time domain measurement\added{ and is in agreement with the expectation calculated in Sec.~\ref{sec:exp_setup}}. A dephasing time $T_2=100\,\mathrm{ns}$ is estimated from the intrinsic peak broadening in spectroscopy\cite{Braumueller2013} and confirmed by related time domain data. The Hamiltonian underlying the simulation reads
\begin{eqnarray}
\hat H^{sim} & = & \hat{H} + \hbar A_q \sum_{i,j} \frac{g_{ij}}{g_{01}}|i\rangle\langle j| \cos\omega_{\mu w} t\\
& \equiv & \hat{H} + \hbar A_q \sum_{i,j} \hat g _{ij} ^x\cos\omega_{\mu w} t,
\label{eq:Hsim}
\end{eqnarray}
which is the generalized Jaynes-Cummings Hamiltonian (\ref{eq:Hmjc}) extended by the applied continuous microwave drive of amplitude $A_q$ and coupling operators $\hat g _{ij} ^x$. For the $x$-coupling of the microwave drive to the anharmonic circuit, we apply the given adaptation of the coupling matrix to account for the circuit's non-linearity. The time evolution of the Liouvillian which is calculated from the master equation of the considered Hamiltonian takes a steady state after roughly \replaced{$500\,\mathrm{ns}\gtrsim T_1, 2\pi/\kappa$}{$500\,\mathrm{ns}\gtrsim T_1, \kappa/2\pi$}, which enables to extract the equilibrium occupation number $\langle \hat n \rangle$ of the anharmonic circuit.
The readout tone of the VNA, applied in the experiment to measure the dispersive shift, is not included in the simulation which is justified by the fact that it is at least three orders of magnitude smaller in power than the applied microwave tones and off-resonant with the multi-level circuit.

To illustrate the good agreement between experimental data and theoretical model, several horizontal slices of the measured power spectrum are compared to simulated data in Fig.~\ref{fig:power_spectrum}(d)-(g). The emergence of clear multi-photon transition signatures for increasing drive power is visible in theory and experiment. The good agreement in position and shape of the measured signatures verifies the validity of the parameters provided to the simulation. The deviation in peak position for the five-photon transition (g) is explained by a frequency dependence of a geometric experimental parameter that is not included in the simulation.

%-----------------------------------------------------------------------------

\section{\label{sec:Rabi_AT}{Rabi sidebands of multi-photon coupled levels}}

%\subsection{Experimental strategy}

Having studied and characterized the multi-photon transitions in the investigated anharmonic circuit in detail in the previous section, we are now able to achieve a dressing of our system involving up to five energy levels by applying a single microwave drive tone. The drive amplitude is denoted as $\Omega_R'$, stressing that the respective Rabi frequency $\Omega_R$ of multi-photon coupled levels can be inferred from it. This is achieved by taking into account the dipole moment of individual transitions and the damping of the readout resonator\cite{Baur09}.
Choosing a drive frequency $\omega_{\mu w} ^d$ that matches one of the observed multi-photon transitions $\frac 1 j \omega_{0j} ^m$ leads to an degeneracy of the driven level $|j\rangle$ and the ground state $|0\rangle$ in energy when regarding the system in the frame rotating with the drive. In this picture we study the drive power dependent dressing of involved levels leading to Rabi sidebands.

The modified level structure in the frame rotating with the drive frequency is probed with a weak microwave tone of frequency $\omega_{\mu w} ^p$ by dynamical coupling of dressed levels $|n'\rangle$. We observe an effective change in the expectation value of the circuit's occupation number $\langle\hat n \rangle$ when the probe frequency matches one of the emerging Rabi sideband transitions.

Applying the unitary transformation
\begin{equation}
\hat U=\exp \left\{i\omega_{\mu w} ^d t\left(\sum_j j|j\rangle\langle j| + \hat a ^{\dagger} \hat a\right)\right\}
\label{eq:Ut}
\end{equation}
on the Hamiltonian (\ref{eq:Hmjc}) yields the level structure
\begin{equation}
\tilde H=\sum_j \left(\omega_j - j\omega_{\mu w} ^d\right)|j\rangle\langle j|
\label{eq:Ht}
\end{equation}
of the undressed anharmonic multi-level system regarded in the frame rotating with $\omega_{\mu w} ^d$. Since the experiment is carried out in the presence of the readout resonator, Eq.~(\ref{eq:Ht}) yields correct results for the transitions in the rotating frame when inserting measured frequencies $\omega_j =\omega_j ^m$.

%\subsection{Master equation simulation including microwave pump and probe tones}

The Hamiltonian used for the master equation simulation of this pump-probe experiment reads
\begin{eqnarray}
\hat H & = & \hbar \sum_j \omega_j |j\rangle \langle j| + \hbar \omega_r \hat a ^{\dagger} \hat a + \hbar \sum_{i,j} g_{01} \hat g_{ij}^x \left( \hat a ^{\dagger} + \hat a \right)\nonumber\\
& & + \hbar \Omega_R' \sum_{i,j} \hat g _{ij} ^x\cos\omega_{\mu w} ^d t+ \hbar A_q ^p \sum_{i,j} \hat g _{ij} ^x\cos\omega_{\mu w} ^p t,
\label{eq:Hsim_03}
\end{eqnarray}
given by extension of Eq.~(\ref{eq:Hsim}) with a microwave drive tone of amplitude $\Omega_R'$ and frequency $\omega_{\mu w} ^d$. The probe tone is written with an amplitude $A_q ^{p}$ and frequency $\omega_{\mu w} ^p$. For better visibility of the features in simulated data, \added{the} provided coherence parameters are enhanced by a factor of two.

\subsection{Two-photon pumping of the $|2\rangle$-level}
\begin{figure} %++++++++++++++++++++++++++++++++++++++FIGURE++++++++++++++++++++++++++++++++++++++++++
\includegraphics{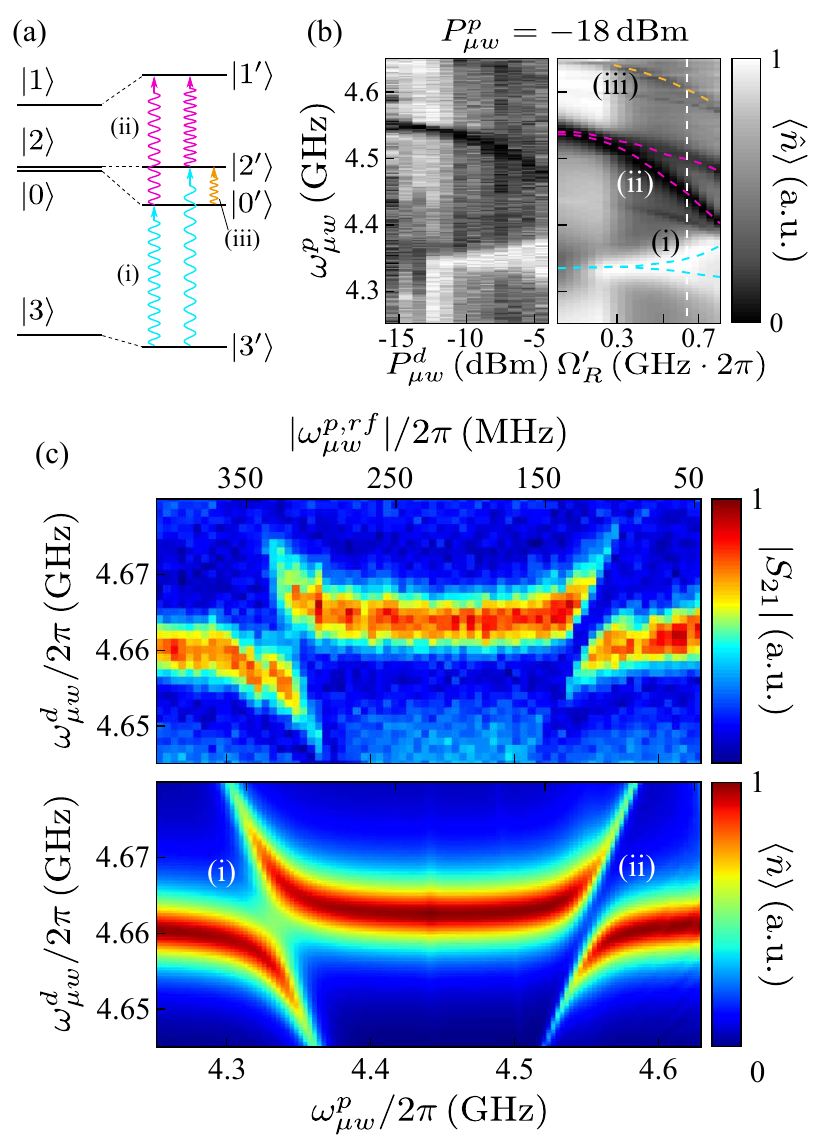}
\caption{(a) Energy diagram in the frame rotating with $\frac 1 2 \omega_{02}$. Switching on the microwave drive leads to a dressing of states, denoted as $|n'\rangle$. Transitions corresponding to the traced Rabi sidebands in (b) are indicated. (b) Measured (left) and simulated (right) Rabi sidebands. A microwave drive of frequency $\omega_{\mu w}^d\added{/2\pi}=4.662\,\mathrm{GHz}$ is applied. Columns are normalized, respectively, in both plots. While $\langle\hat n\rangle$ is reduced for the $|0'\rangle\leftrightarrow|1'\rangle$ sideband transition (ii), occupation is effectively increased for the $|0'\rangle\leftrightarrow|3'\rangle$ sideband (i). (c) Measured and simulated data of the associated Autler-Townes doublets for $P_{\mu w} ^d=-14\,\mathrm{dBm}$ and $\Omega_R'=0.2\added{/2\pi}\,\mathrm{GHz}$. The probe frequency $\omega_{\mu w} ^{p,rf}$ in the rotating frame is given on the above horizontal axis.}
\label{fig:2ATsidebands}
\end{figure}

Setting the microwave drive to $\omega_{\mu w} ^d \added{/2\pi}=\frac 1 2 \,\omega_{02} ^m\added{/2\pi}=4.662\,\mathrm{GHz}$, Eq.~(\ref{eq:Ht}) yields the undressed energy levels in the rotating frame
\begin{equation}
\begin{array}{ccrc}
\omega_3 ^{rf}/2\pi & = & (-)330\,\mathrm{MHz} & (i),\\
\omega_1 ^{rf}/2\pi & = & 104\,\mathrm{MHz} & (ii).
\label{eq:02transitions}
\end{array}
\end{equation}
The related energy diagram of the circuit regarded in the rotating frame is depicted in Fig.~\ref{fig:2ATsidebands}(a).

In Fig.~\ref{fig:2ATsidebands}(b), we observe the recently reported\cite{KoshinoPRL13} Rabi sideband transitions (ii), (iii) between the first three transmon levels traced in pink (ii) and orange (iii). These features, corresponding to the $|0'\rangle \leftrightarrow |2'\rangle$ transition (iii) and the $|0'\rangle \leftrightarrow |1'\rangle$ transition (ii) in the dressed system cause an effective decrease (dark signature) in occupation number $\langle\hat n'\rangle $ relative to the driven saturation state $|n'\rangle =\frac 1 2 |0'\rangle + \frac 1 2 |2'\rangle$.
Taking into account higher levels, the three-photon sideband transition $|0'\rangle \leftrightarrow |3'\rangle$ traced in light blue is visible as a bright feature since $\langle\hat n'\rangle $ effectively increases as the $|3'\rangle$-level is populated. Bending and splitting of the Rabi sidebands in Fig.~\ref{fig:2ATsidebands}(b) account for a stronger dressing with increasing drive amplitude $\Omega_R'$. The repelling of levels is most pronounced for directly coupled levels $|0'\rangle$, $|2'\rangle$, since they coincide in energy in the rotating frame without any applied drive. Dressing of levels $|1\rangle$ and $|3\rangle$ is less pronounced, as the energetic distance to the nearest level is larger, see Eq.~(\ref{eq:02transitions}). Due to the increased dressing, indicated sidebands (i), (ii) approach each other, in agreement with the simple picture shown in Fig.~\ref{fig:2ATsidebands}(a). It is important to note that a higher probe frequency $\omega_{\mu w} ^p$ in the laboratory frame corresponds to a smaller frequency $\omega_{\mu w} ^{p,rf}$ in the rotating frame.

Tuning the drive frequency into the vicinity of the resonant condition allows to observe an Autler-Townes doublet whenever the applied probe tone matches a transition in the dressed system. Figure~\ref{fig:2ATsidebands}(c) shows the Autler-Townes doublets of the two dominant transitions (i), (ii) involving the dressed zero-level $|0'\rangle$, as indicated in the energy diagram. Since level saturation $|n'\rangle =\frac 1 2 |0'\rangle +\frac 1 2 |2'\rangle$ is not fully achieved for the small drive amplitude $\Omega_R'$ chosen in the simulation, the driven level $|2'\rangle$ is only weakly populated and therefore a dynamical coupling to other levels, inducing an Autler-Townes splitting, is weak. For very small drive amplitudes, the doublets are observable at frequencies calculated from Eq.~(\ref{eq:02transitions}) for the undressed system. To observe the Autler-Townes splittings in experiment at a reasonable signal-to-noise ratio we apply a microwave drive which is roughly by a factor of two higher in amplitude as compared to simulated data. Therefore, the observed splittings approach each other in Fig.~\ref{fig:2ATsidebands}(c), upper plot, in accordance with the bending of respective Rabi sidebands shown in Fig.~\ref{fig:2ATsidebands}(b).

\subsection{Three-photon pumping of the $|3\rangle$-level}
\begin{figure} %++++++++++++++++++++++++++++++++++++++FIGURE++++++++++++++++++++++++++++++++++++++++++
\includegraphics{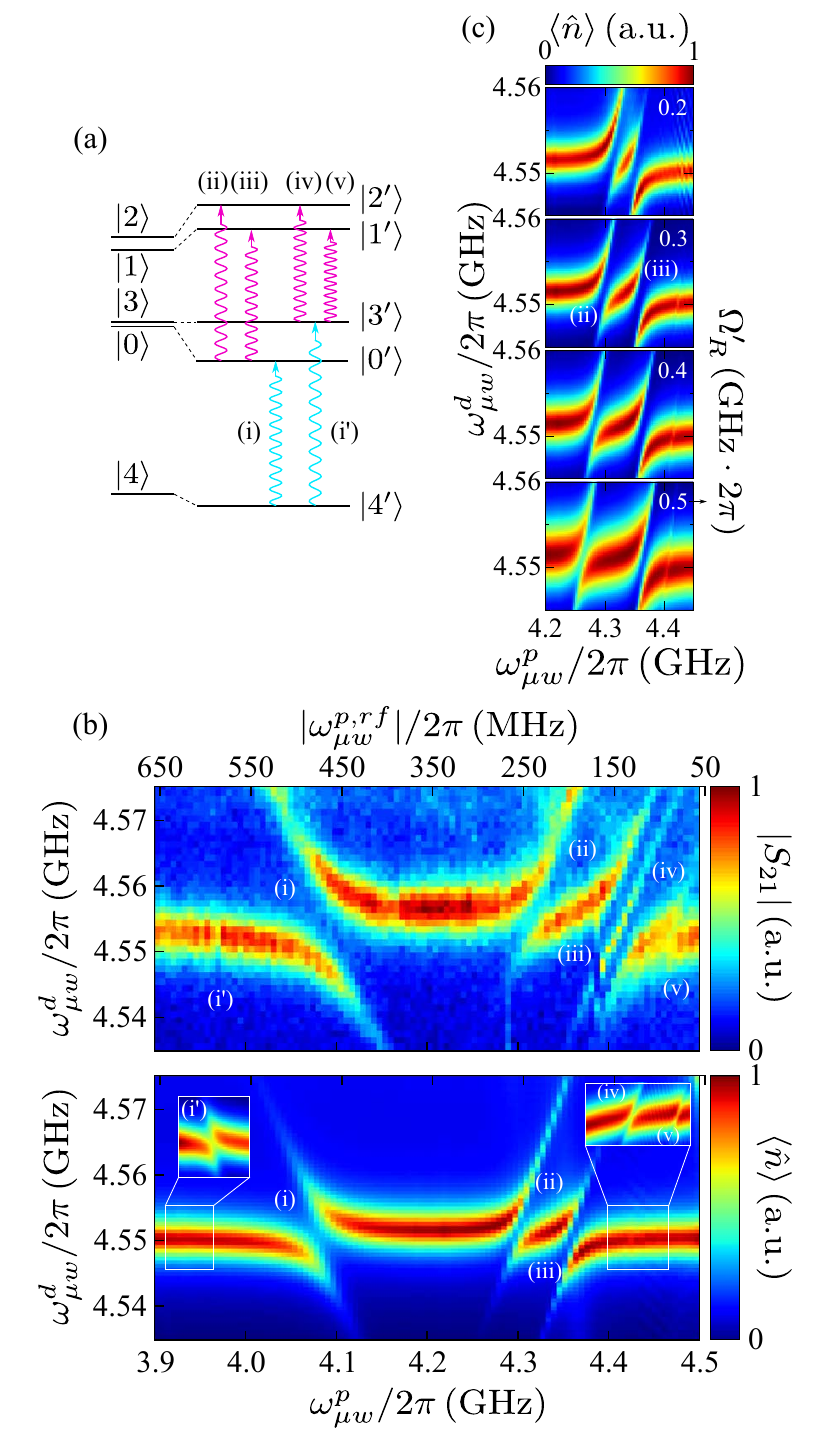}
\caption{(a) Energy diagram in the frame rotating with $\frac 1 3 \,\omega_{03}$ without (left) and with (right) microwave drive. (b) Measured and simulated data of the Autler-Townes splittings corresponding to the Rabi sideband transitions indicated in (a). A drive power $P_{\mu w} ^d=-5\,\mathrm{dBm}$ and a probe power $P_{\mu w} ^p=-13\,\mathrm{dBm}$ are applied in the measurement. Simulation parameters are $\Omega_R'=0.3/2\pi\,\mathrm{GHz}$ and $A_{q} ^p=0.1/2\pi\,\mathrm{GHz}$. (c) Enlarged view of the simulated Autler-Townes triplet (ii), (iii) dependent on the Rabi frequency $\Omega_R'$. Repulsion of the transitions for stronger dressing is demonstrated.}
\label{fig:3ATsidebands}
\end{figure}

In a similar manner we achieve dressing of the anharmonic circuit involving up to five levels by means of three-photon pumping. Transforming into the frame rotating with frequency $\omega_{\mu w} ^d \added{/2\pi}=\frac 1 3 \,\omega_{03} ^m \added{/2\pi}=4.552\,\mathrm{GHz}$, levels $|0\rangle$ and $|3\rangle$ are degenerate. Applying a microwave drive matching this three-photon transition in frequency lifts the degeneracy and induces a dressing of states, as depicted in the energy diagram of Fig.~\ref{fig:3ATsidebands}(a). The transitions in the undressed system regarded in the frame rotating with the applied drive frequency $\frac 1 3 \omega_{03}$ are calculated from Eq.~(\ref{eq:Ht}) as
\begin{equation}
\begin{array}{ccrc}
\omega_4 ^{rf}/2\pi & = & (-)466\,\mathrm{MHz} & (i),\\
\omega_1 ^{rf}/2\pi & = & 214\,\mathrm{MHz} & (ii),\\
\omega_2 ^{rf}/2\pi & = & 220\,\mathrm{MHz} & (iii).
\label{eq:03transitions}
\end{array}
\end{equation}
The five stable levels corresponding to localized states of our circuit are therewith dressed by a single microwave tone.
Probing the circuit with a weak microwave tone allows to observe the generalized multi-photon Autler-Townes doublets in the three-photon pumped dressed system. Measured data and the related master equation simulations are presented in Fig.~\ref{fig:3ATsidebands}(b). The dominant doublets again arise from the single-photon processes involving the dressed zero-level $|0'\rangle$, (i)-(iii), since level saturation $|n'\rangle =\frac 1 2 |0'\rangle + \frac 1 2 |3'\rangle$ is not achieved for small drive amplitudes. They coincide in position with the transition frequencies calculated from Eq.~(\ref{eq:03transitions}) for the undressed system. In measured data, fainter features corresponding to dynamical coupling with the $|3'\rangle$-level are visible since higher microwave amplitudes are chosen to guarantee good visibility above the noise level. This explains small deviations in the exact position of the doublets between measurement and simulation. Features (i'), (iv) and (v) are emphasized in the inset of the simulation by increasing the probe amplitude. Transitions related to the observed Autler-Townes splittings are indicated in the energy diagram given in Fig.~\ref{fig:3ATsidebands}(a).
Since energy levels $|1\rangle$, $|2\rangle$ are very close in the undressed system, their strong mutual repelling is observable by a distinct double splitting even for small drive amplitudes. 
Figure~\ref{fig:3ATsidebands}(c) shows simulated data of the Autler-Townes double-splitting for varying drive amplitudes $\Omega_R'$. The dependence of the dressing on the Rabi frequency is clearly confirmed by the dispersion of the Autler-Townes doublets for increasing $\Omega_R'$. The pronounced repulsion of levels  $|1'\rangle$, $|2'\rangle$ provides direct evidence for the comprehensive dressing of the considered five-level subsystem.
As the dressing increases, the frequency of the $|1'\rangle \leftrightarrow |3'\rangle$ transition (v) can be observed to be constant (not shown), which is explained by the close proximity of levels $|1'\rangle$, $|2'\rangle$ to the pumped level $|3'\rangle$, likewise exerting a repulsive effect.

%-------------------------

\section{\label{sec:conclusion}{Conclusion}}

We have investigated a superconducting anharmonic many-level circuit coupled to a harmonic readout resonator. We observe higher-order multi-photon transitions up to the fifth excited state via virtual energy levels of our circuit. The back-action of these multi-photon excitations on our readout device is analyzed quantitatively and demonstrated to be in accordance with theoretical expectation. For the investigation of a multi-level quantum circuit it is essential to tune the readout frequency close to the inflection point of the resonator's Lorentzian to minimize necessary corrections induced by the non-linearity.
By applying a strong microwave pump tone we achieve multi-photon dressing of our circuit involving up to five energy levels. Rabi sidebands are observable by probing the dressed system with a weak microwave tone. The expected dependence of the Rabi frequencies on the drive power is demonstrated. Associated higher-order Autler-Townes doublets are observed for multi-photon pumping levels $|2\rangle$ and $|3\rangle$, appearing at frequencies matching the expectation from analysis of the circuit Hamiltonian in the rotating frame. Experimental results are in agreement with master equation simulations taking into account the full Hilbert space of the investigated anharmonic circuit.

%-----------------------------------------------------------------------------

\begin{acknowledgments}
This work was supported by Deutsche Forschungsgemeinschaft (DFG) and the State of Baden-W\"urttemberg through the DFG-Center for Functional Nanostructures (CFN).
This work was also supported in part by the Ministry for Education and Science of the Russian Federation Grant 11.G34.31.0062, in the framework of the Increase Competitiveness Program of the National University of Science and Technology MISIS under contract number K2-2014-025 and by Deutscher Akademischer Austauschdienst (DAAD) with project number 57056114.
J.B. would like to acknowledge financial support by the Helmholtz International Research School for Teratronics (HIRST). 
\end{acknowledgments}

%-----------------------------------------------------------------------------

%\nocite{*}   %show all entries of bib-file
\bibliography{multi-photon_s_2c}  %bibliography via BibTeX.


\end{document}